\documentclass[a4paper,11pt]{article}

\usepackage{jcappub}
\usepackage[utf8]{inputenc}
\usepackage{enumitem}
\usepackage{graphicx}
\usepackage{hyperref}
\usepackage{mathtools}
\usepackage{comment}
\usepackage{amsmath}
\usepackage{ amssymb }
\usepackage{lipsum} 
\usepackage{float}
\usepackage{cleveref}
\usepackage[caption=false]{subfig}
\usepackage{placeins}
\usepackage{xcolor}
\usepackage{float}
\usepackage[normalem]{ulem}

\usepackage{natbib}
\usepackage{comment}

\def\be{\begin{equation}}
\def\ee{\end{equation}}
\def\bea{\begin{eqnarray}}
\def\eea{\end{eqnarray}}

\def\s{{\rm s}}
\def\cm{{\rm cm}}
\def\kpc{{\rm kpc}}
\def\mev{{\rm MeV}}
\def\gev{{\rm GeV}}
\def\erg{{\rm erg}}

\title{Applying Simulation-Based Inference to Spectral and Spatial Information from the Galactic Center Gamma-Ray Excess}

\author[a]{Katharena Christy,}
\author[b]{Eric J.~Baxter,}
\author[a]{Jason Kumar}

\affiliation[a]{\mbox{Department of Physics \& Astronomy, University of Hawai`i, Honolulu, HI 96822, USA}}
\affiliation[b]{\mbox{Institute for Astronomy, University of Hawai`i, 2680 Woodlawn Drive, Honolulu, HI 96822, USA}}

\emailAdd{chri3448@hawaii.edu}

\abstract{The two most favored explanations of the \textit{Fermi} Galactic Center gamma-ray excess (GCE) are millisecond pulsars and self annihilation of the smooth dark matter halo of the galaxy. 
 In order to distinguish between these possibilities, we would like to optimally use all information in the available data, including photon direction and energy information.
To date, analyses of the GCE have generally treated directional and energy information separately, or have ignored one or the other completely.
Here, we develop a method for analyzing the GCE that relies on simulation-based
inference with neural posterior models to jointly analyze photon directional and spectral information while correctly accounting for the spatial and energy resolution of the telescope, here assumed to be the {\it Fermi} Large Area Telescope (LAT). 
Our results also have implications for analyses of the diffuse gamma-ray background, which we discuss.}  

\begin{document}


\maketitle


\section{Introduction}

In spite of its observed abundance, the nature of dark matter (DM)  
remains elusive.
If DM is made of weakly interacting massive particles (WIMPs) with some non-zero cross section for annihilation to Standard Model particles, a gamma-ray flux may be observable from regions of high DM density. In 2009, 
an excess of gamma rays from the Galactic Center (GC) was discovered using data from the {\it Fermi} Large Area Telescope 
(LAT)~\cite{Goodenough:2009gk,Hooper:2010mq,Abazajian:2012pn,Fermi-LAT:2015sau}.  
This Galactic Center excess (GCE) is roughly consistent with the expected spectral and spatial distributions of WIMP DM~(see, for example,~\cite{Goodenough:2009gk,Calore:2014xka}).
However, a definitive identification of the excess with dark matter annihilation requires ruling out possible astrophysical (i.e. non-DM) explanations.   Some alternative explanations of the GCE have gained traction, such as populations of unresolved point sources. The most viable point source candidates are millisecond pulsars (MSPs)~\cite{Abazajian:2010zy}.  There has been a great deal of study regarding how well DM and MSPs can explain the {\it Fermi} data, and how they might be distinguished.  The stakes are high, since if the dark matter explanation can be confirmed, this would have major implications for our understanding of dark matter and physics beyond the Standard Model.

Due to the uncertainties in MSP and DM models, efforts to distinguish between these possibilities have been 
evolving rapidly.
Given known backgrounds, the spectral shape of the GCE is broadly consistent with the gamma-ray spectrum of both MSPs and of some dark matter models (for example, dark matter with a mass $m_{\chi} \sim 30-50~\gev$ annihilating to the $b \bar b$ final state~\cite{Goodenough:2009gk,Calore:2014xka}. 
Spatially, although some analyses~\cite{DiMauro:2021raz,Cholis:2021rpp,McDermott:2022zmq,Zhong:2024vyi} have found that  the excess is generally consistent 
with DM annihilation in a halo close to Navarro-Frenk-White (NFW), other recent analyses find that the GCE is spatially correlated with the stellar mass distribution of the galactic bulge, favoring the MSP interpretation~\cite{Macias:2016nev,Bartels:2017vsx,Storm_2017,Macias:2019omb,Ploeg_2021,Pohl:2022nnd,Song:2024iup}.  
In particular, Ref.~\cite{Song:2024iup} showed that an MSP signal following the stellar bulge distribution of Ref.~\cite{Coleman:2019kax} was significantly preferred (in the sense of Bayesian evidence) to that of a dark matter signal, with Ref.~\cite{Zhong:2024vyi} finding 
the likelihoods of these models to be similar (but making no attempt at Bayesian model comparison).
The last word on this subject may not be written, and a detailed study of the spectrum and  morphology of the GCE is not the 
goal of this work.
      
Beyond the overall morphology of the GCE, another form of spatial information that has been considered as a means of distinguishing between DM and  MSP explanations is the photon counts-in-pixels distribution (CPD), which quantifies the probability, $P(C)$, of observing $C$ photons in a pixel.\footnote{In cases where this probability distribution is the same for every pixel on the sky, the CPD is just proportional to the histogram of photon counts in pixels.}  Since the DM signal comes from the combined flux of many volumes of space, each with a very small probability of producing a DM annihilation photon, the  CPD from DM annihilation is expected to be Poisson (with slowly varying mean across the sky).  The MSP signal, on the other hand, originates from a population of point sources.  If some of these sources are just below the point source detection threshold of the LAT, then a single pulsar may produce multiple photons detected by the LAT, leading to a non-Poisson CPD~\cite{Lee:2015fea}.
Thus, even if the \textit{expectation value} of the dark matter and pulsar signals were the same in each pixel, the two may produce different fluctuations around this expectation.  
The CPD (without energy information) has been previously used to analyze the GCE, but the results remain inconclusive, 
as mismodelling of the background can significantly bias the analysis~\cite{Leane:2019xiy,Buschmann:2020adf,Leane:2020nmi,Zechlin_2016}. 
Moroever, depending on the details of the MSP luminosity function, the MSP CPD can be very close to Poisson, and thus hard to distinguish from that of the DM \cite{Dinsmore_2022}.

As we have said, our goal is not to definitively describe the spectrum and morphology of the GCE, but to describe a 
methodology which can improve analyses of the GC by utilizing more relevant information.
In order to improve constraints on the DM and MSP contributions to the GCE, we would like to include multiple sources of information --- spatial and spectral --- from the data.    If this information were completely decoupled, one could simply analyze the photon directions and energies separately, and combine the resultant posteriors. In general, though, the spectral and spatial information are coupled: the DM and MSP signals may differ both spatially and spectrally from each other. To fully and correctly extract this information, then, one must jointly analyze the spatial and spectral data.  This is the goal of the mock analysis that we present below.

Moreover,  additional information  may be contained in \textit{correlations} between the spectral and spatial data.  For instance, the spectrum of MSPs is expected to vary from pulsar to pulsar while the DM spectrum is constant across the sky.  
If the GCE arises due to gamma-ray emission from a relatively small population of bright pulsars just below 
the detection threshold, then each pulsar would produce several photons observed by Fermi-LAT, with a spectrum which varies from pixel to 
pixel.  Such spectral variations would be absent if the GCE arises from DM annihilation.

In addition to offering improved constraints on the DM and MSP models, a joint analysis of energy and spatial information from LAT observations is also necessary because the point spread function (PSF) of \textit{Fermi} is energy-dependent.  This induces a coupling between the spatial and energy information which can only be correctly modeled if both are considered simultaneously.

Effort has been made toward joint spatial and spectral GCE analysis~\cite{Boyarsky:2010dr,Abazajian:2012pn,Abazajian:2014fta}, including recent work using adaptive-constrained template fitting \cite{Storm_2017, Calore_2021}.
In this work, however, we introduce a technique for jointly analyzing spatial and energy information (as well as their correlations) in the LAT data that relies on simulation-based inference (SBI).  This analysis naturally includes information in the CPD.  We will study the improvements in constraining power that can be obtained through such an analysis, as well as the biases that may be introduced by \textit{not} including the coupling between spatial and energy information.\footnote{Note that there may be additional sources of information in the data beyond spatial and spectral that could be used to distinguish between DM and MSPs and that we do not consider here, such as photon timing information \cite{Baxter:2023}.}

The spatial and spectral data can be encapsulated in a $N_{\rm pix}$ pixels map of the sky across $N_E$ different energy bins, yielding a data vector $\vec{d}$ with length $N_E N_{\rm pix}$.  As long as the pixel size is comparable to the beam size and the energy binning is comparable to the energy resolution, then we expect minimal information to be lost in energy/spatial binning. 
If one can compute a likelihood for the data, $P(\vec{d}|\theta)$, given model parameters $\theta$, 
then standard Bayesian tools such as MCMC can be used to constrain the parameters of the DM and MSP models, and to assess the preference in the data for DM over MSPs. This approach, sometimes under the name of non-Poisson template fitting (NPTF) has been previously applied to analyses of the GCE \cite{Leane:2019xiy,Buschmann:2020adf,Leane:2020nmi} and the diffuse gamma-ray sky \cite{Lee:2009, Baxter:2010, Runburg:2021} when $N_E = 1$. 

However, the combinatorics of computing the exact likelihood quickly become intractable as the number of pixels, energy bins, and model components increase (see, for example,~\cite{Baxter:2021tui}).  For this reason, previous analyses of the GCE relying on the exact likelihood approach have 
often been restricted to a single energy bin.  Moreover, the telescope PSF complicates the likelihood calculation because it couples the photon counts in nearby pixels. 

An alternative to the exact likelihood calculation is to obtain an approximate posterior through simulation, an approach known as 
simulation-based inference (SBI) or likelihood-free inference (LFI). As drawing from the relevant distributions to simulate data avoids the computationally intractable combinatorial sums in the calculation of the likelihoood, it is much easier to simulate photon count maps than to compute their likelihood.  
SBI therefore provides a powerful approach for analyzing the gamma-ray data  that can potentially lead to improved constraints by allowing for the incorporation of more information from the data~\cite{Baxter:2021tui}.  

SBI has been previously applied to the analysis of gamma-ray data by several authors (see, for example,~\cite{List:2020mzd,List:2021aer}).  
\citet{Mishra-Sharma:2021oxe} recently applied SBI to analysis of the GCE to distinguish between DM and MSP explanations.  That work used a similar SBI approach to the one we employ here.   However, the analysis of Ref.~\cite{Mishra-Sharma:2021oxe} does not incorporate energy information, relying instead on only the spatial information (which implicitly includes information in the CPD).   In Ref.~\cite{Baxter:2021tui} we applied the SBI  method of Approximate Bayesian Computation (ABC) to the diffuse, extragalactic gamma-ray sky, including both energy and CPD information.  In ABC, the posterior is estimated by using a distance metric to  compare the data to simulations generated at different points in parameter space.  Values of the parameters that yield simulations within some distance threshold of the true data will provide an approximate sample from the posterior in the limit that the distance threshold is small.  We found that constraints on contributors to  the diffuse extragalactic background could be significantly improved through the addition of energy information, enabled by SBI.

Here, we present a proof of concept application of the SBI method of Neural Posterior Estimation (NPE), to a mock analysis of the GCE, including spatial and spectral models of backgrounds, foregrounds, DM annihilation, and pulsar populations.  Our analysis naturally includes spatial and energy information, and their correlations.  Our main goals are to (1) demonstrate how the inclusion of both spatial and spectral information can break degeneracies between the DM and MSP models, (2) assess the relative information content of spatial/spectral data and their correlations, and (3) demonstrate the impact of correctly treating the instrumental PSF and spectral dispersion.  

An important difference between the present work and previous applications of SBI to gamma-ray data concerns the nature of our simulations.  Here, we simulate individual photons with assigned directions and energies, and only later produce pixelized photon count maps as a form of data compression.  Previous analyses have instead worked entirely with pixelized maps.  The disadvantage of starting from a pixelized approach is that  effects like the telescope PSF and energy resolution cannot be treated entirely correctly.  With only pixelized maps, it is not possible to correctly incorporate the impact of the PSF on the simulations since the distribution of photon positions within the pixel is not known.\footnote{Note, though, that the average impact of the PSF can be modelled, see e.g.~Ref.~\cite{Malyshev:2011}.}   This may be a small effect if the pixel size is small compared to the PSF.  However, it can have a significant impact on the non-Poisson tails of the CPD, precisely the regime that carries information about the presence of unresolved point sources like MSPs. 
Because we generate individual photon data, we can exactly simulate the impact of the PSF and energy dispersion by perturbing the directions and energies of individual photons.

Our approach of generating individual photons is more computationally demanding than generating photon counts maps since it requires building a full 3D simulation of the distribution of MSPs around the Galactic Center (see further discussion in \S\ref{sec:photon_generation}).  However, because the total number of photons and MSPs is small (at most about $10^5$), this additional computation is feasible. Below, we will explore the impact of our exact treatment of the PSF and energy dispersion on our results.

The plan of the paper is as follows.  In Section~\ref{sec:GammaRaySourceModels}, we describe our models for potential sources of gamma rays from the GC, including dark matter 
annihilation, MSPs, and diffuse backgrounds.  In \S\ref{sec:photon_generation} we describe our methodology for generating photons from the source models.  In \S\ref{sec:sbi}, we describe our application of simulation-based inference to train a neural posterior model that relates data to posterior.  We present our main results in \S\ref{sec:results}, and conclude in \S\ref{sec:discussion}.

\section{Gamma-ray Source Models}
\label{sec:GammaRaySourceModels}

We consider three possible sources for gamma-rays from the GC: dark matter annihilation, 
MSPs, and the diffuse astrophysical background, which originates primarily from the interactions of cosmic rays with the galactic matter and radiation fields.  For each source, we model both its spatial and spectral distribution.  We will assume that point sources above the detection threshold of Fermi-LAT have been detected and removed.

\subsection{Dark Matter}
We will assume that dark matter annihilates from an $s$-wave state, and that the dark matter 
halo has the form of a generalized NFW (gNFW) profile:
\begin{eqnarray}
\rho_{\rm \chi} (r) = \rho_s (r/r_s)^{-\gamma} [1+(r/r_s)]^{\gamma -3},
\end{eqnarray}
with $\gamma = 1.2$, $\rho_s = 10^7 M_\odot \kpc^{-3}$, and $r_s = 20~\kpc$.  This scenario has been found to be a reasonable fit to the GCE morphology (see, for example, Ref.~\cite{Daylan:2014rsa}), and so we will adopt it as our benchmark.  

The flux due to dark matter annihilation in some direction $\theta$ from the GC is then given by 
\bea
\Phi_{\rm \chi} \equiv \frac{d^2 \phi}{dE d\Omega} (\theta) &=& \frac{\langle \sigma v \rangle_0}{8\pi m_\chi^2} \frac{dN_{\gamma}}{dE} 
\int d\ell~\rho_{\rm \chi}^2 (r(\ell, \theta)) ,
\eea
where $\phi$ is the flux in photons per time, $m_\chi$ is the dark matter mass, $\langle \sigma v \rangle_0$ is the 
$s$-wave dark matter annihilation 
cross section, and $dN_{\gamma}/dE$ is the photon spectrum due to dark matter annihilation.  Here, 
$\ell$ is the distance along the line of sight and $r(\ell, \theta) = \sqrt{\ell^2 + D^2 - 2 \ell D \cos \theta}$, 
where $D$ is the distance to the GC.   
We will assume that dark matter annihilates with 100\% branching fraction to the 
$b \bar b$ final state, as this scenario has been found to be a reasonable fit to the spectrum of the GCE if $m \sim 30-50~\gev$.  Note, though, that the methods we develop are completely general, and could just as well be applied to other dark matter annihilation models.
We determine the gamma ray spectrum for this model using the results of Ref.~\cite{Cirelli:2010xx}. Example dark matter spectra for different particle masses are shown in Fig.~\ref{fig:spectra} (left panel).

\subsection{Pulsars}
\label{sec:pulsars}

Following Ref.~\cite{Mishra-Sharma:2021oxe}, we include 
two contributions to the total pulsar population: one which follows the GC excess, and another which follows the density of the Galactic disk.  We refer to these as `core' and `disk' pulsars, respectively.  For the core pulsars, we assume that their spatial 
distribution is proportional to $\rho_{\rm \chi}^2 (r)$, meaning that the spatial distribution of photons emitted by the core pulsars is the same as from the dark matter.
Thus, the spatial distribution of the core pulsars is given by
\begin{eqnarray}
    \frac{dN_{\rm core}}{dV} = \frac{A_{\rm core}}{V_0} \frac{\rho_{\rm \chi}^2(r)}{\rho^2_s},
\end{eqnarray}
where  $A_{\rm core}$ is a free parameter and $V_0 \equiv \int_{\Omega} d\Omega \int d\ell~\ell^2\rho_{\rm \chi}^2 (r) / \rho^2_s$ is a 
normalization chosen so that $A_{\rm core}$ is the expected number of GC pulsars in the region of interest for our analysis (see \S\ref{sec:roi}).

Recent work has pointed out that pulsar distribution tends to follow the stellar distribution~\cite{Macias:2016nev,Bartels:2017vsx,Macias:2019omb,Song:2024iup}, rather than that of an gNFW-distributed dark matter signal.  
However, we choose a pulsar spatial distribution which is the same as that of a dark matter signal in order to better illustrate the ability 
of our analysis to distinguish between these sources using combined spectral and spatial information.  Our choice 
minimizes the information contained in the mock data GCE morphology alone, allowing us to more clearly see the gain obtained 
by analyzing spectral and spatial information jointly.

The disk pulsars are assumed to be distributed 
according to:
\bea
\frac{dN_{\rm disk}}{dV}=A_{\rm disk}\exp(-r^2/(2\sigma_r^2))\exp(-|z|/z_0),
\eea
where $z$ is the height above the plane of the galaxy, $r$ is the cylindrical radial coordinate with origin at the Galactic Center, and $A_{\rm disk}$ is a free parameter. Ref.~\cite{Ploeg_2020}
finds a best-fit model (their model {\it A1}) that has
$\sigma_r = 4.5_{-0.4}^{+0.5}~\kpc$ and $z_0 = 0.71_{-0.09}^{+0.11}~\kpc$.   We will adopt those values here.  
However, as we will describe later, we will mask photons from the galactic plane.  After masking, we will find that the expected 
contribution of photons from the disk pulsars is very small, leading to essentially no constraint on this population.  Thus, while we will marginalize over the presence of a possible disk pulsar population in our analysis, we do not present constraints on this population in the main text.

In addition to specifying the spatial distributions of the pulsars, we must also specify their luminosity distributions.  We take the luminosity function to be~\citep{Dinsmore_2022}:
\bea
\frac{1}{N} \frac{dN}{dL} = {\frac{(1-n_1)(1-n_2)}{L_b(n_1-n_2)}}
\begin{cases}
(\frac{L}{L_{b}})^{-n_1}, L<L_{b}\\
(\frac{L}{L_{b}})^{-n_2}, L>L_{b}
\end{cases}.
\eea
We consider a  benchmark luminosity model with 
$n_1=-0.66$, $n_2=18.2$, $L_b = 2.5 \times 10^{34} \erg / \s$.  
This benchmark 
was obtained in Ref.~\cite{Lee:2015fea} as the best fit to NPTF in the inner galaxy, without 
masking the galactic plane (masking the plane changes the best fit slightly).
For this model,  $\langle L \rangle \sim 0.66 L_b$, and the number of pulsars 
(resolved or unresolved) producing the GCE is $A_{core } = 637$. 
The flux from GCE pulsars is then $F \sim 1.8 \times 10^{-9} \erg~\cm^{-2} \s^{-1}$, which is in good agreement with the 
literature~\cite{Dinsmore_2022}.  In this case, a relatively small number of pulsars each produce a 
relatively large number of gamma rays seen by Fermi, resulting in pronounced non-Poisson fluctuations in the 
photon count distribution.

\begin{figure}
    \includegraphics*[width=0.45\textwidth]{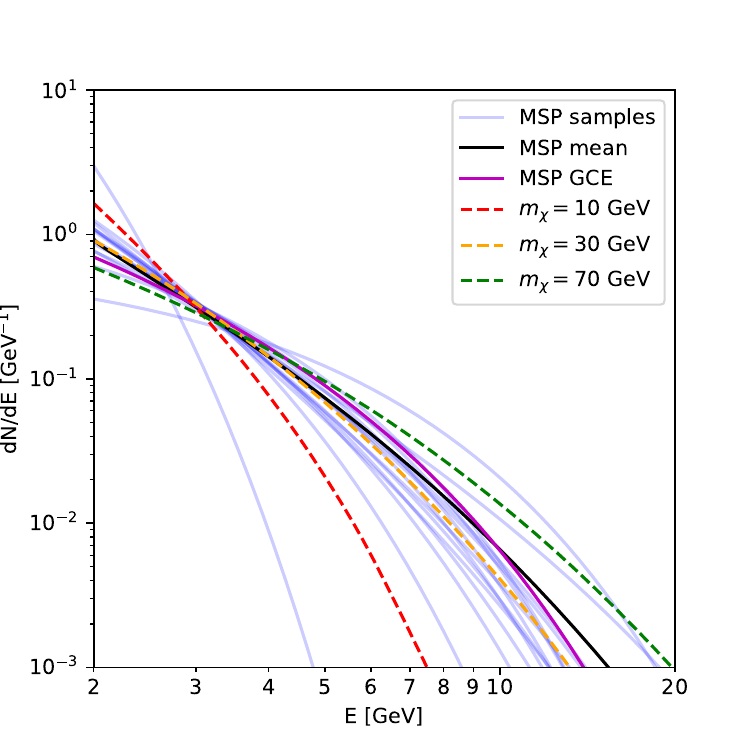}
     \includegraphics*[width=0.6\textwidth]{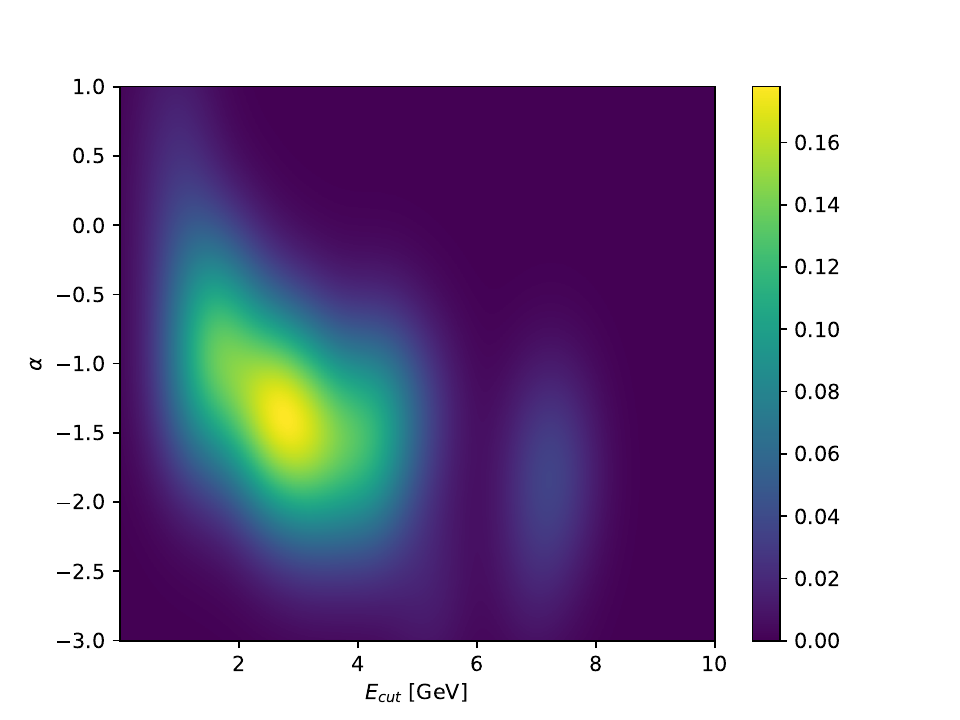}
    \caption{{\it Left}: sample spectra of the various model components. Spectra for dark matter annihilation to $b\Bar{b}$ 
    are shown for $m_{\chi} = $ 10 (red dashed), 30 (orange dashed) and 70 (green dashed) GeV. 
    Also shown are 30 sample MSP spectra (solid light blue) drawn from the distribution in the right panel, as well as 
    the average of spectra drawn from this distribution (solid black).  A best fit to the GCE (solid magenta) is also 
    shown.    
    {\it Right}: Gaussian kernel density estimation of the distribution of MSP spectral parameters ($E_{\rm cut},\alpha$), derived from known galactic MSPs fit in Ref.~\cite{Cholis:2014noa}. }
    \label{fig:spectra}
\end{figure}

For the pulsar spectrum, we follow the results 
of Ref.~\cite{Cholis:2014noa}, which fit the spectra of 61 pulsars observed by {\it Fermi} to a power-law $\times$ exponential model:
\bea
\frac{1}{N_{\gamma}} \frac{dN_{\gamma}}{dE} &\propto&  \frac{E^\alpha}{E_{\rm cut}^{1+\alpha}} e^{-E / E_{\rm cut}} .
\label{eq:MSPspectrum}
\eea
Since the values of $\alpha$ and $E_{\rm cut}$ vary from pulsar 
to pulsar, we will assume that  these spectral parameters are drawn from some underlying distribution, which can be inferred 
from the spread in spectra of the 61 pulsars studied in 
\cite{Cholis:2014noa}. 
We estimate this distribution by applying kernel density estimation~\citep{10.1214/aoms/1177728190,10.1214/aoms/1177704472}
to that data, 
assuming a Gaussian kernel.  The resulting distribution is plotted in Figure~\ref{fig:spectra} (right panel).

In Fig.~\ref{fig:spectra} (left), we plot 30 MSP spectra drawn from this distribution.  We also 
plot 
the average spectrum obtained by drawing many sources from this distribution (denoted ``mean").  
A fit of this spectrum to the functional form in 
Eq.~\ref{eq:MSPspectrum} yields \cite{Cholis:2014noa} 
\bea
\alpha &=& -1.57 {}^{+0.001}_{-0.002} ,
\nonumber\\
E_{\rm cut} &=& 3.78 ^{+0.15}_{-0.08} \gev .
\label{eq:MSPstacked}
\eea
In Fig.~\ref{fig:spectra} (left), we also plot a spectrum of the form given in Eq.~\ref{eq:MSPspectrum} with 
$\alpha = 0.945$, $E_{\rm cut} = 2.53$ (denoted ``GCE"),
which was found in Ref.~\cite{Calore:2014xka} to provide the best fit to the GC excess.
Fig.~\ref{fig:spectra} indicates 
that the level of variability in individual MSP spectra is significant.

Note that pulsars in the bulge may have different luminosities and spectral parameters than those in the disk 
due to distinct star formation histories. In this case, one may not be able to extrapolate from observations of disk pulsars to the luminosity and spectral functions of MSPs producing the GCE.
However, these differences are likely not statistically significant~\cite{Ploeg_2020}, and we will not consider this possibility further.
More generally, there is still uncertainty regarding both the MSP spectrum and the dark matter spectrum.  The latter arises because 
we do not know a priori either what the dark matter particle mass is, or what the branching fractions are to all possible final states.
But our goal is to determine the ability of a joint analysis of spatial and energy information to distinguish between 
a DM or MSP origin of the GCE.  For this purpose, it is not so important that the pulsar or dark matter spectra be correct as that they 
be sufficiently similar to each (i.e., to illustrate that the analysis works even in the maximally difficult scenario).  
From Fig.~\ref{fig:spectra} it is clear that pulsar spectra are generally quite similar to that produced 
by dark matter annihilation to $b \bar b$ with $m_{\chi} \sim 30~\gev$.
But we will also consider a case in which the dark matter spectrum is taken to be exactly the same as the 
average pulsar spectrum, representing the case of maximal similarity of MSP and DM spectral information (in the absence of 
spatial information).  We will see that, compared to that worst case, any differences between the DM annihilation spectrum and the average 
MSP spectrum will only serve to improve our ability to constrain the origin of the GCE.

On the other hand, if the variability of MSP spectra (that is, how much spectra produced by MSPs in 
a pixel can differ from the average) is significantly smaller than our estimates, then the ability of this method to 
distinguish between an MSP or DM origin of the GCE will decrease. The variability of the MSP spectra may be less 
than our estimates, either because MSPs in the GCE may have inherently less variability than those considered in \cite{Cholis:2014noa}, 
or because there are many faint MSPs in a pixel, resulting a smoothing out of the gamma-ray spectrum produced by the collection 
of MSPs.  We will consider this case in the Appendix.

\subsection{Diffuse astrophysical background}

Our model for the diffuse astrophysical background, $\Phi_{\rm diffuse}$ is given by
\begin{eqnarray}
    \Phi_{\rm diffuse} = \Phi_{\rm iso} + \Phi_{\rm aniso} + \Phi_{\rm bubbles},
\end{eqnarray}
where $\Phi_{\rm iso}$, $\Phi_{\rm aniso}$, and $\Phi_{\rm bubbles}$ are respectively: an isotropic background which may include e.g. extragalactic contributions, an
anisotropic background that traces the structure of the galaxy, and the Fermi bubbles \cite{fermibubbles}.  We describe the spatial and spectral models for these components in more detail below.  For each source, we build a model for the flux $\Phi$, which is a function of both energy and direction on the sky.  These flux models are computed  on \texttt{healpix}\footnote{\url{https://healpix.jpl.nasa.gov/}} maps with $N_{\rm side} = 64$ (corresponding to $N_{\rm pix} = 49152$) and in $N_E = 10$ log-spaced energy bins from $2 - 100~\gev$.

\subsubsection{Isotropic emission}

Our model for the flux from isotropic diffuse sources on the sky, $\Phi_{\rm iso}$, is derived from the Fermi isotropic background model.\footnote{\url{https://fermi.gsfc.nasa.gov/ssc/data/access/lat/BackgroundModels.html}}  This spectral model includes contributions from diffuse and unresolved extragalactic sources, and cosmic rays that are misclassified as gamma rays.  The model is derived from fits to existing Fermi data.

\subsubsection{Diffuse aniostropic background model}

Our model for the flux from the anisotropic diffuse background, $\Phi_{\rm aniso}$, is taken from  the Model O of Ref.~\cite{Buschmann:2020adf}.\footnote{\url{https://github.com/nickrodd/FermiDiffuse-ModelO}.}  
Model O includes two contributions to the galactic gamma-ray background: 
\begin{enumerate}[label=(\roman*)]
\item interactions of cosmic rays with galactic gas.   Two processes are relevant here: first, inelastic scattering of cosmic ray protons with gas, producing pions which subsequently decay and produce $\gamma$s, and second, Bremsstrahlung radiation from cosmic ray electrons interacting with the gas.
\item inverse Compton scattering of cosmic ray electrons with the interstellar radiation fields.    
\end{enumerate}
Model O includes templates for both (i) and (ii) which we utilize here, setting the amplitudes to those obtained in Ref.~\cite{Buschmann:2020adf}.  

\subsubsection{Fermi bubbles}

Our model for the flux from the Fermi bubbles, $\Phi_{\rm bubbles}$, uses the spatial template from {\it DSpace@MIT},\footnote{\url{https://dspace.mit.edu/handle/1721.1/105492}} 
while the spectrum is taken from Ref.~\cite{Su:2010qj}.

\section{Generating simulated data}
\label{sec:photon_generation}

We now describe the generation of mock photons from the source models. 

For the purposes of our mock analysis, we assume an exposure of $1.2 \times 10^{11}~\cm^2\s$, similar to Ref.~\cite{Mishra-Sharma:2021oxe}, and roughly the current exposure of Fermi LAT. 
With this exposure, the energy arriving at {\it Fermi} as GCE photons is $\sim {\cal O}(100)~\erg 
\sim {\cal O}(10^5)~\gev$.

\subsection{Generating photons from Poisson sources}

Some of the sources considered above, namely dark matter annihilation and the diffuse backgrounds, will generate photon counts in each pixel that are Poisson distributed with an expectation value that varies slowly across the sky.  

The expected number of photons in each pixel from such a source for an energy bin specified by $(E_{\rm min}, E_{\rm max})$ is given by 
\begin{eqnarray}
\langle C \rangle = {\rm Exp} \times \Omega_{\rm pixel} \int_{E_{\rm min}}^{E_{\rm max}} dE \,\Phi,
\end{eqnarray}
where ${\rm Exp}$ is the exposure in units of area $\times$ time, and $\Omega_{\rm pixel}$ is the solid angle of the pixel. We sample from a Poisson distribution with mean $\langle C \rangle$ to determine the number of photons from the source in this pixel and energy bin.  
In order to correctly treat the effect of the PSF, we generate diffuse photons using high-resolution pixels with size $\Omega_{\rm pixel}$ which is much smaller 
than the pixel size eventually used for the analysis.  Each photon thus generated has a true direction known to much greater precision than the 
size of the pixel used in analysis.

\subsection{Generating photons from MSPs}

Unlike the Poisson sources, photons from MSPs arise from a small number of point sources, each of which can produce multiple photons.  To generate these photons, we first draw a realization of the positions of  MSPs around the galactic center.  We begin by assuming that the number of MSPs in any differential volume element is 
Poisson-distributed about a mean given by the integral of the number density, $dN/dV$ over the volume.  
The number of pulsars in a given volume is then assigned by drawing from this distribution.  For each pulsar, 
a luminosity is drawn from the corresponding luminosity distribution, $dN/dL$ and the parameters of the spectrum
are also drawn from the estimated distribution described in \S\ref{sec:pulsars}, yielding a unique location, spectrum and luminosity for each pulsar.  
  
The actual number of photons from the pulsar is then drawn from a Poisson distribution whose mean is the expected number of photons given the pulsar's luminosity and distance, and the assumed exposure. Each photon is assigned an energy by randomly drawing from the spectral distribution of that pulsar (the spectrum of each pulsar varies, as noted previously).  Finally, the photons from 
the pulsar are assigned to the same direction on the sky as the pulsar  relative to Earth.

\subsection{Application of the PSF}

As each photon has a true energy and direction, it is now possible to exactly account for the 
angular and energy resolution of {\it Fermi} LAT.  
We use the {\it Fermi} LAT point spread function model detailed in the {\it Fermitools Cicerone}.\footnote{\url{https://fermi.gsfc.nasa.gov/ssc/data/analysis/documentation/Cicerone/Cicerone_LAT_IRFs/IRF_PSF.html}}
The amplitude of the angular deviation of every incident photon direction due to the PSF is given by 
\bea
\delta p = x \cdot S_p(E),
\label{eq:AngRes}
\eea
where $x$ is a stochastic factor and $S_p(E)$ is the energy dependent scale factor given by

\bea
S_p(E) = \sqrt{\left[ c_0 \left( \frac{E}{100 \: \mev} \right) ^{-\beta} \right] ^2 + c_1^2}.
\eea

Here, $c_0,$ $c_1,$ and $\beta$ are determined by the event class and event type. We assume use of the cleanest data class of P8R3\_ULTRACLEANVETO, and the event type corresponding to the most accurate quartile of reconstructed direction, PSF3. The distribution of the factor $x$ is given by 
\begin{equation}
P(x, f, \sigma, \gamma) = f_{\rm core}K(x, \sigma_{\rm core}, \gamma_{\rm core})+ 
(1-f_{\rm core})K(x, \sigma_{\rm tail}, \gamma_{\rm tail}),
\end{equation}
where $K$ is the 
King function
\bea
K(x,\sigma, \gamma) = \frac{1}{2\pi\sigma^2}\left( 1-\frac{1}{\gamma}\right)\left[ 1+\frac{1}{2\gamma}\frac{x^2}{\sigma^2} \right]^{-\gamma}.
\eea
The parameters $f$, $\sigma$, and $\gamma$ are determined by the event class and type, as well as the energy and inclination angle. Although there is some dependence of the PSF on inclination angle, we assume normal incidence throughout.  This is justified by our chosen event class and type being those  closest to this approximation. 
The so-called Fisheye Effect causes further deviation of the PSF from azimuthal symmetry due to lower reconstruction efficiency at incidence angles further from the LAT boresight. However, this effect is minimized at normal incidence and is subsequently ignored in our analysis.  Each photon is perturbed by a $\delta p$ randomly drawn from the distribution above.

\subsection{Application of energy dispersion}

We model The Fermi LAT energy dispersion following the {\it Fermitools Cicerone},\footnote{\url{https://fermi.gsfc.nasa.gov/ssc/data/analysis/documentation/Cicerone/Cicerone_LAT_IRFs/IRF_E_dispersion.html}\label{footnote:irf}} 
which gives  the deviation in reconstructed photon energy as
\begin{equation}
\delta E = E \cdot x \cdot S_D(E, \theta),
\end{equation}
where the scale factor $S_D(E, \theta)$ is given by
\begin{equation}
S_D(E, \theta) = c_0(\log E)^2 + c_1(\cos\theta)^2 + c_2\log E  
+c_3\cos\theta + c_4\log E\cos\theta + c_5.
\end{equation}
The parameters $c_{0-5}$ are determined by the event class and type, taken to be P8R3\_ULTRACLEANVETO and PSF3, respectively, as above. The distribution of the stochastic factor $x$ is

\begin{equation}
D(x) = F \cdot g(x; \sigma_1, k_1, b_1, p_1) + (1-F) \cdot g(x; \sigma_2, k_2, b_2, p_2),
\end{equation}

where the base function $g$ is given by

\begin{equation}
g(x;\sigma,k,b,p) = \frac{p}{\sigma\Gamma(1/p)}\frac{k}{1+k^2}
\times \begin{cases}
    \exp(-\frac{k^p}{\sigma^p}|x-b|^p, & \: x-b \geq 0)\\
    \exp(-\frac{1}{k^p\sigma^p}|x-b|^p, & \: x-b < 0)
\end{cases}.
\end{equation}
All parameters are determined by the assumed event class, event type, energy, and angle of incidence,
and are taken from the {\it Fermi} LAT instrument response function (see footnote~\ref{footnote:irf}).  
The energy of each photon is perturbed by a $\delta E$ randomly drawn from distribution above.

\subsection{Restriction to region of interest}
\label{sec:roi}
We restrict our analysis to a region-of-interest  
(ROI) that is within $10^\circ$, of the Galactic Center.  We exclude the region with galactic latitude $|b| < 2^\circ$ in order to reduce the sensitivity to modeling of the galactic disk.  
We consider photons in the energy range $2 - 100~\gev$.  In this energy range, 
the angular resolution is $< 0.4^\circ$ and the energy resolution is $< 10\%$ (68\% containment)~\cite{Fermi-LAT:2009ihh}.
Photons are generated within a larger region (within $15^\circ$ of the Galactic Center, 
masking $|b| < 1^\circ$), in order to 
allow the effects of the PSF to be properly accounted for.  
Photons are generated in 
a larger energy range ($1 - 150~\gev$), in order to ensure that the effects of 
the energy resolution can be properly accounted for.

\section{Application of SBI}
\label{sec:sbi}

\subsection{Summary statistic}

Our raw data are very high dimensional: we have a measured direction on the sky and energy for every photon.  In order to achieve better results with our neural posterior-based SBI analysis, we compress the data into various summary statistics.  

\begin{figure}
    \includegraphics*[width=0.47\textwidth]{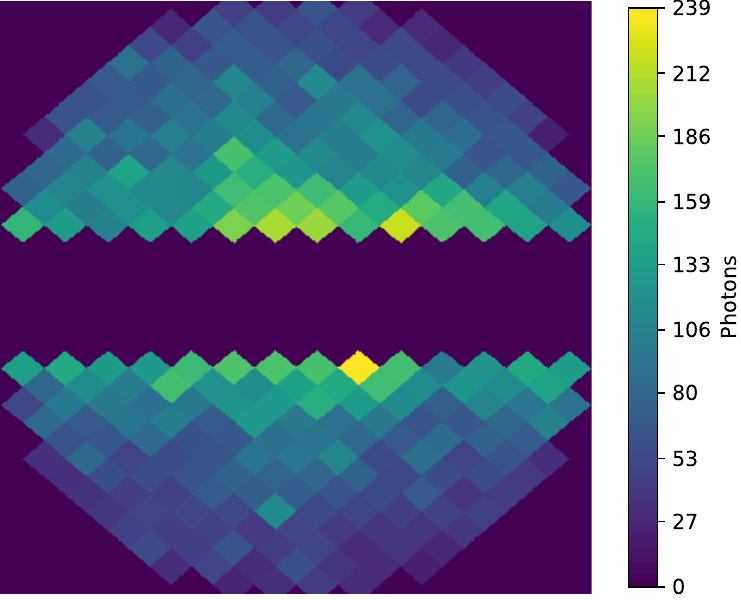}
    \includegraphics*[width=0.52\textwidth]{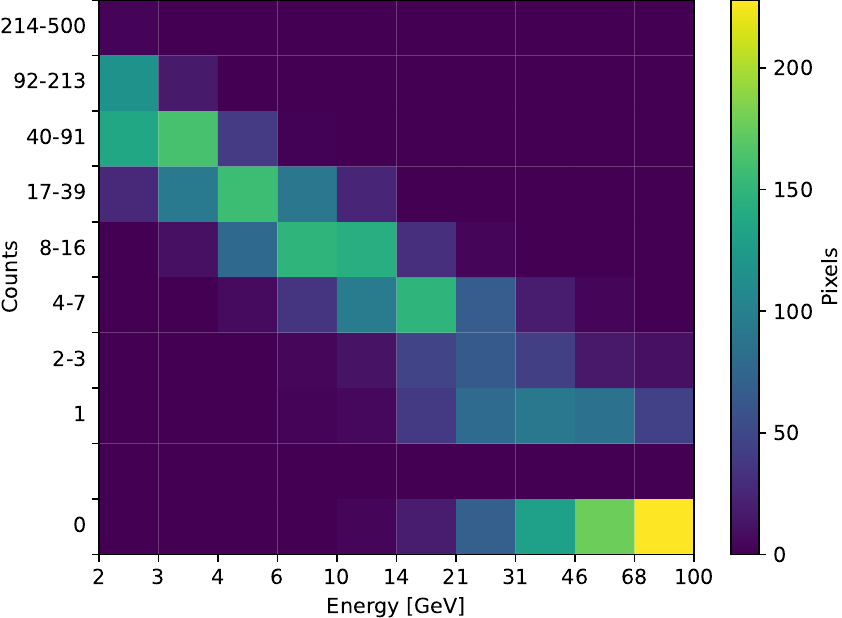}
    \caption{{\it Left}: example photon count map of the ROI used in our analysis.  For illustration, we show only photons in the energy range of 2 \gev $<$ E $<$ 2.96 \gev. Colors indicate the number of photons in a pixel.  The central region is masked, as discussed in the text.  This map corresponds to a model with $(A_{\rm core},$ $m_\chi,$ $\langle \sigma v \rangle_0) = (319,$ $30~\gev,$ $10^{-26}~\cm^3 \s^{-1})$.
    {\it Right}: example of the energy-dependent histogram summary statistic used for the SBI analysis. Columns correspond to different energies, while rows correspond to different counts bins.
    }
    \label{fig:HPmap}
\end{figure}

First, the data are compressed into pixelized counts maps in  energy bins (an example is shown in the left panel of Fig.~\ref{fig:HPmap}).  To this end, we use the same \texttt{healpix} pixelization scheme and energy binning as described above.  In principle, these pixelized maps could be used as the final summary to which SBI is applied.  Indeed, this is similar to the approach taken by Ref.~\cite{Michra-Sharma:2021}, although their analysis used a single energy bin and trained a neural network to further compress the data before training their neural posterior model.  Here, we instead  compress the energy-dependent counts maps to the energy-dependent histogram of photon counts.  This is just the histogram of photon counts in our pixelized maps at each energy bin.  An example of this summary is shown in Fig.~\ref{fig:HPmap} (right).  This summary preserves the spatial information inherent in the photon CPD, but removes information in large scale variations in the CPD. 

Because our analysis is restricted to a region away from the galactic plane over which the expected photon counts do not vary by huge amounts, we do not expect very much information to be lost by taking this approach.  More importantly, though, our choice of summary statistic enables us to directly explore the discriminatory power of the spectral and CPD information, as well as their correlations.  This is one of the primary goals of the present analysis.

\subsection{Neural posterior model}

We use the method of Neural Posterior Estimation (NPE) developed in Ref.~\cite{NPE} to train a model for the posterior on the model parameters given simulations of the data at different points in parameter space.  This trained model can then be used to estimate the posterior given a future (real or simulated) datastet.  The basic idea of NPE is to posit a general parametric model for the posterior, in this case represented by a neural network.  It was shown in Ref.~\cite{NPE} that if this model is trained by maximizing the probability of the parameters  given the simulated data, the result can be turned into an estimate of the true posterior by weighting the learned function by a combination of the prior and the distribution used to propose parameters for the simulations.   The neural network architecture that we use here is a stack of masked autoregressive flows, following the defaults in the \texttt{sbi} package  \citep{sbicode}.  Training is performed using $10^5$ simulations.  We have tested that our results are robust to variations in the training set size.

\section{Results}
\label{sec:results}

We now present the results of our SBI analysis of simulated data.  The simulated data used in the mock analysis are generated in the same manner as the simulations used to train the SBI model.  Of course, real data from Fermi may differ in important ways from our simulated data, which could lead to biased parameter constraints in an actual data application.  We postpone an investigation of this issue to future work, as our focus here is mainly on understanding the information content of the data, and how it can be extracted.  For this purpose, ensuring that our simulator perfectly matches real data is less important.

\subsection{Improvements from joint energy and directional information}

\begin{figure}
    \includegraphics*[width=\textwidth/2]{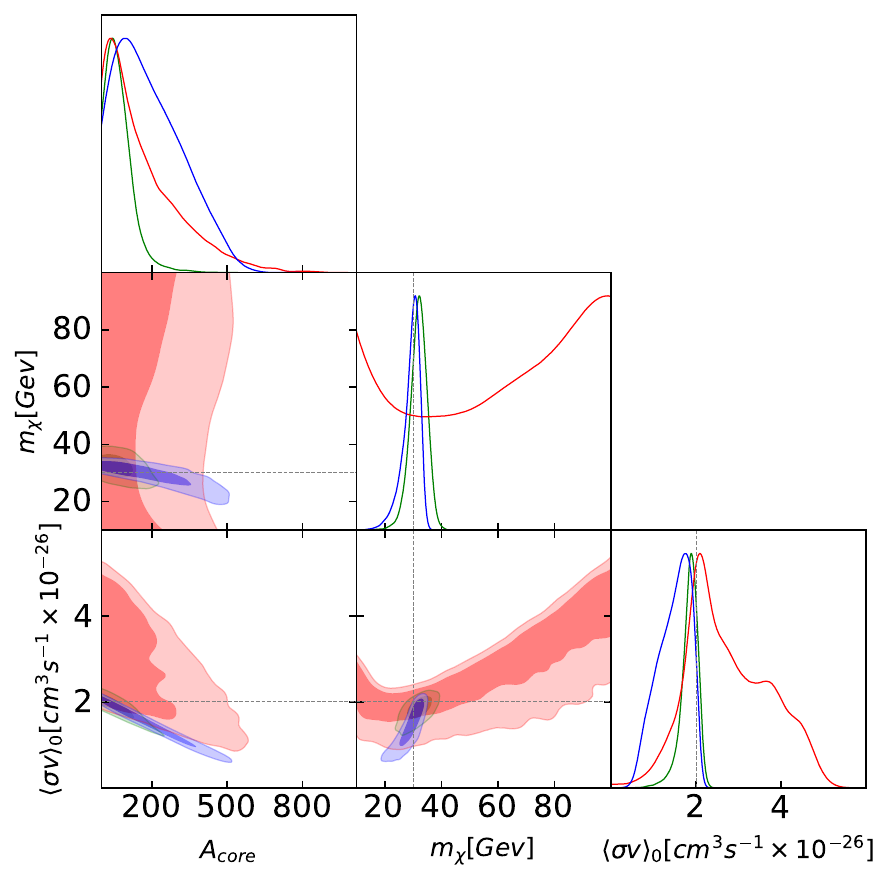}
    \includegraphics*[width=\textwidth/2]{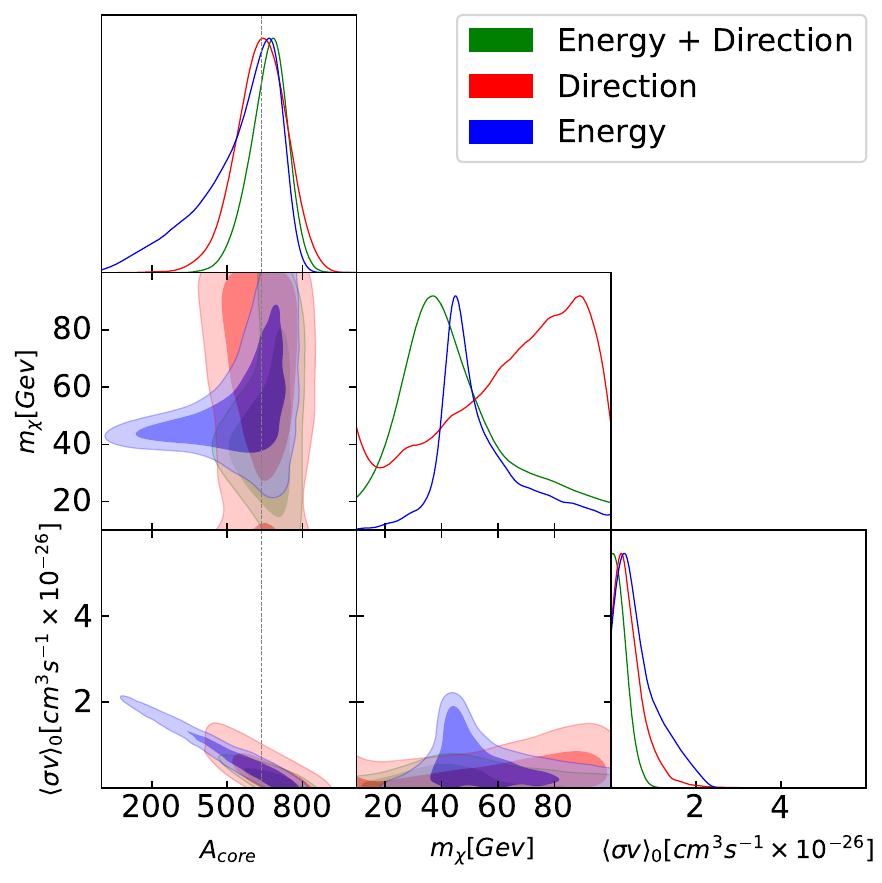} \\
    \begin{center}
     \includegraphics*[width=\textwidth/2]{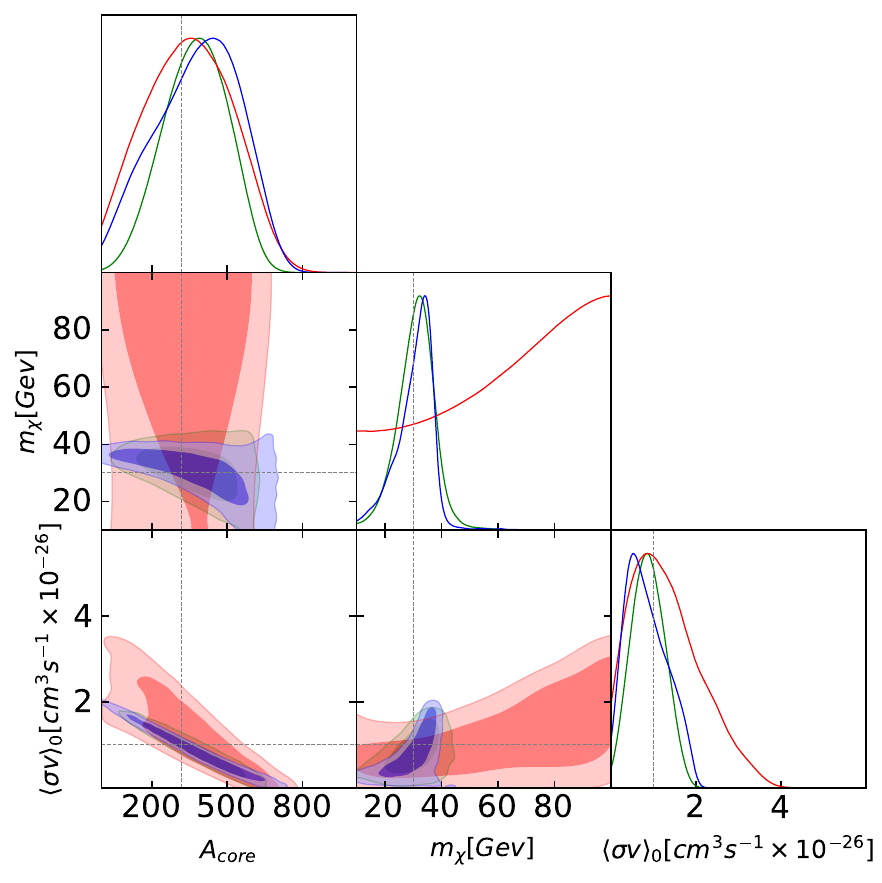}
    \end{center}
    \caption{Posteriors on the model parameters from our mock analyses. 
 The blue contours represent the case when only energy information (i.e.~the spectrum) is used, the red contours correspond to using only directional information in the form of the energy-independent CPD, and the green contours correspond to using both energy and directional information via the energy-dependent CPD. The mock data are the same in each of these cases, and are generated using a true model in which the source of the GCE is only dark matter (top left), only pulsars (top right), or in which the MSPs and DM contribute roughly equal numbers of photons to the GCE (bottom).  In all cases, the true model is indicated by the grey dotted lines.  
}
     \label{fig:contours}
\end{figure}

In Figure~\ref{fig:contours}, we compare the parameter constraints which are obtained when using 
joint energy and directional information with those obtained when only directional information in the form of the CPD, or only energy information (that is, the  spectrum) are used. 
We consider the cases in which the 
true model has only a dark matter contribution to the GCE (top left panel), 
only a MSP contribution (top right panel), and the case in which both 
dark matter and MSPs contribute about equally to the GCE (bottom panel). 
The dark matter contribution is taken to arise from a $30~\gev$ particle annihilating to 
$b \bar b$.  Green contours represent parameter constraints obtained with the energy-dependent  
CPD, while the red contours represent constraints obtained with the 
energy-independent CPD.  The blue contours are constraints obtained 
using the number of total counts in each energy bin, and  contain no 
spatial information.
The dark (light) contours correspond to 
$1\sigma$ ($2\sigma$) credibility.
 
Even if no energy information is used we find that we are generally able to recover parameters 
which are consistent with the true model.\footnote{Note that, in the absence of energy information, the dark matter mass and cross section are highly degenerate.  However, the key point here is that we can distinguish between dark matter and MSPs, as evidenced by the fact that the DM annihilation cross section and $A_{\rm core}$ are not fully degenerate.}  Since we have set the pulsar spatial distribution to match that expected for the dark matter annihilation signal (when viewed on the sky), our ability to distinguish between DM and MSP explanations of the GCE without energy information arises from the information in the CPD.  The inclusion of energy information 
significantly reduces the size of the contours.  In particular, if the true model of the GCE consists 
of either DM-only or MSPs-only, then the use of energy information allows one to confirm at $2\sigma$ 
credibility, the presence of that source.  The improvement obtained from using energy information is particularly 
striking, given that the DM and pulsar contributions have very similar spectra by design.  Notably, although the 
parameter constraints obtained from the energy-dependent counts-in-pixels histogram are the tightest, 
even the constraints obtained from the spectrum only are sufficient to rule out a DM (MSP)-only origin 
of the GCE, if the true origin is in fact MSPs (DM).  
This indicates that, despite the similarity between the DM and MSP spectra, 
they are nevertheless easily distinguishable statistically, given the large number of photons comprising the 
GCE.  

However, this analysis is somewhat unrealistic, as we have not included the effects of background mismodelling, 
or of modifications to the dark matter annihilation spectrum which might arise from a more complicated final state.  
It is unlikely that one can realistically model the backgrounds and signals with such precision that one can 
statistically rule out either DM or MSPs as the origin of the GCE from spectral information alone, given the 
general similarity of these spectra.

To quantify how well one can constrain the origin of the GCE, even if the 
spectra of DM and pulsar models are indistinguishable, we will now consider a case in which the photon spectrum arising 
from dark matter annihilation is taken to be exactly the same as the average pulsar spectrum.  In this case, 
one can only use the total counts in  energy bins to discriminate between a DM and MSP origin for the GCE through  
non-Poisson fluctuations in the number of photons in an energy bin, driven by the fact that each MSP produces a different 
photon spectrum.

\begin{figure}
    \includegraphics*[width=\textwidth/2]{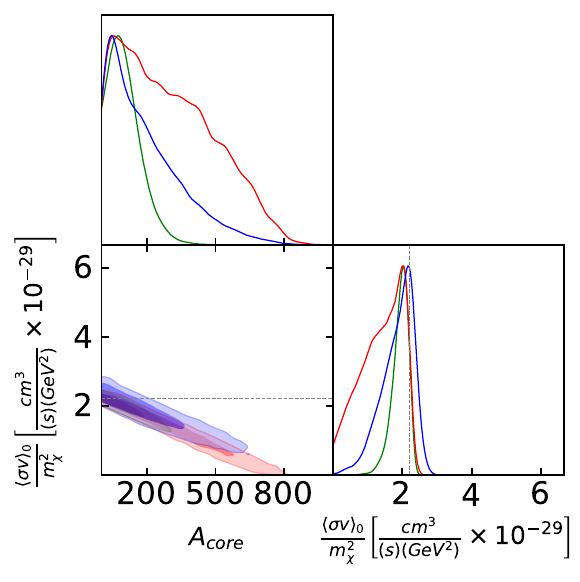}
    \includegraphics*[width=\textwidth/2]{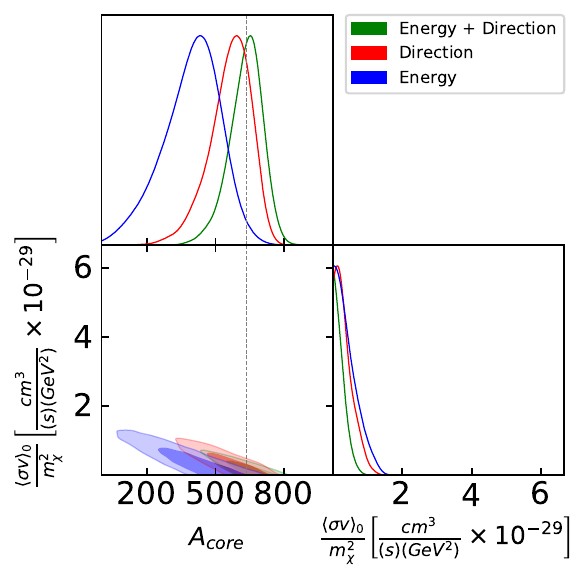}
    \caption{Similar to top panels of Fig.~\ref{fig:contours}, but in the case for which the dark matter spectrum 
    is taken to be identical to the average MSP spectrum.   }
    \label{fig:red}
\end{figure}

We illustrate this in Figure~\ref{fig:red}, which shows parameter 
constraints in the case where the GCE arises entirely from DM (left panel) and MSPs (right panel).  In each case, 
the dark matter spectrum is taken to be the average of the MSP spectrum.  As a result, the only parameter 
of the DM model is the normalization, $\langle \sigma v \rangle_0 / m_\chi^2$.  Again, the green contours 
are obtained from the energy-dependent counts-in-pixels histogram, while the red contours are obtained from 
the energy-independent counts-in-pixels histogram, and the blue contours are obtained from the
energy spectrum alone.  
We see that energy information alone still provides some moderate ability to distinguish 
between MSP and DM origins of the GCE, as the parameter constraints do not show complete degeneracy between the 
two sources.  As the DM spectrum is taken to be the average of MSP spectrum, 
and as no spatial information is used in this case, this discriminating power derives only from the fact that each 
MSP will have a  different spectrum, leading to non-Poissonianity in the photon counts in each energy bin 
for the signal produced by MSPs, as compared to the Poisson-distribution of the photon count in each energy bin 
for the signal produced by DM annihilation.  
Similarly, the use of directional information alone also provides some moderate ability to distinguish 
between MSP and DM origins of the GCE, consistent with the analysis in Fig.~\ref{fig:contours}.
However, it is clear that much more information is available when both energy and spatial 
information are used.  The use of both spatial and energy information allows one to clearly determine the 
presence of either the MSP or DM source, in the cases where either is the exclusive source of the GCE.

As a further illustration, we redo this analysis with a different choice for the pulsar distribution 
and luminosity model in the Appendix. 
For this model, many more MSPs would be needed to generate the GCE.  As a result, the 
photon count distribution from MSPs is much more Poisson-like, and the effect of variations in the MSP spectrum 
from pulsar to pulsar is averaged out, leaving a much smaller net effect.

\subsection{Potential biases in parameter reconstruction}

We can consider the extent to which the parameters we reconstruct are biased in 
relation to the true parameters.  In doing so, we can also consider the extent to 
which correct implementation of the PSF, which requires simulated data genrated at 
the level of individual photons, reduces these biases.  To do so, we generate 100 
mock data samples from a true model in which half of the GCE photons are expected to 
arise from MSPs, and the other half from $30~\gev$ dark matter annihilating to $b \bar b$.
For each of the 100 samples, we find the mean and variance of the 1D posteriors for 
the parameters $A_{\rm core}$, $m_{\chi}$ and $\langle \sigma v \rangle_0$.  We then find the mean and 
variance of the mean of these 100 samples.  

We find that there is some bias in the posterior relative to the 
true parameters,  but this bias is small compared to the size of the $68\%$ credible interval 
of the 1D posteriors obtained from the analysis of any single mock data set.  At that level, 
we see that effect of any residual bias in the posteriors is negligible.

In a similar way, we can also consider the potential biasing effects of  PSF mismodelling that would result from not correctly incorporating the energy dependence of the PSF, as we have done in our analysis.  To do this, we analyze the 
same 100 mock data sets using a posterior model trained on data with an incorrectly modelled PSF which does not 
use the energy information from individual photons  
and instead
computes the PSF with Eq.~\ref{eq:AngRes} using the average expected photon energy over the energy range we consider.  Considering the parameters reconstructed from the 100 mock 
data samples using this incorrectly trained network,  we find that the mean and the variance of the 
mean are biased, in comparison to the analysis with a correctly modelled PSF, by an amount which is small 
compared to the uncertainties in a single analysis.
This result is not surprising, since the size of the pixels we have used is large compared to the 
angular scale of the PSF.

\section{Discussion}
\label{sec:discussion}

We have presented an approach to analyzing gamma-ray data from the GCE that uses both photon directional and energy information, as well as their correlations.  Our analysis naturally incorporates information encoded in the CPD and the photon spectrum, both of which have been previously (but separately) used to analyze the GCE.  Our approach relies on simulation-based inference with neural posterior models.  Applying our analysis to mock data, we find that the combination of photon directional and energy information is particularly powerful for analyzing the GCE and for discriminating between MSP and DM explanations.  At the most basic level, adding energy information to a standard (i.e.~energy-independent) CPD analysis significantly improves constraints on parameters such as the dark matter particle mass.  This is because, in the absence of energy information, the particle mass is completely degenerate with the annihilation cross section.  Additionally, if there are differences between the dark matter and MSP spectra, these could be used to discriminate between the two sources.  For our chosen dark matter model (annihilation to $b\bar{b}$) and MSP model, we find that is indeed the case, given the statistics of current {\it Fermi} data.  However, relying on such spectral differences is risky, since it is unlikely that we know the functional forms of the DM, MSP, and background spectra sufficiently well for this result to be robust.  

However, we find that by \textit{combining} directional and energy information in our joint analysis, we can achieve additional discriminating power that does not rely on spectral differences between the DM and MSPs.  In particular, because the spectrum of each MSP is different (and the differences can be significant), while the spectrum of DM annihilation is fixed, there will be non-Poisson fluctuations in the energy-dependent photon counts distribution that carry additional information.  We find that even if the spectrum of the DM annihilation is identical to the mean MSP spectrum, it is possible to distinguish between the two with our analysis, as a result of these correlated fluctuations.

In addition to including more information, the analysis developed here enables a more exact treatment of observational effects (such as the PSF and energy dispersion) than previous analyses because we apply these effect to individual photons.  We have tailored the details of our analysis to {\it Fermi} LAT, but in principle, these same improvements would carry over to data from any telescope.

We have assumed throughout that photon source distributions have been modelled correctly.  However, there are significant 
uncertainties associated with the diffuse astrophysical background, MSP spectra, luminosity and spatial distributions, 
as well as the dark matter distribution and annihilation channel.  
We have shown that the joint use of 
energy and direction information allows for tighter parameter constraints when all sources are modelled correctly, but it 
would be interesting to investigate if this improvement is robust to some level of source mismodelling.  
As previous studies of the GCE have demonstrated~\citep{Horiuchi:2016zwu,Leane:2019xiy,Leane:2020nmi}, analyses 
of non-Poisson fluctuations can be biased when backgrounds are mismodelled.  At a basic level, it is difficult to distinguish the 
non-Poisson fluctuations of a correctly-modelled background from the Poisson fluctuations of an incorrectly-modelled background.
We expect that our methodology will be more robust against the effects of mismodelling than methods which only use the photon 
counts-in-pixels distribution, because of the joint inclusion of energy information.  For example, if the amplitude or spatial 
distribution of a diffuse background component is mismodelled, then spectral information from photons in a pixel may still 
make it possible to distinguish this scenario from the presence of a signal contribution with a different spectrum.  Moreover, 
the variation of the MSP spectrum from pulsar to pulsar provides a tool which may permit one to distinguish the 
presence of bright unresolved MSPs from a mismodelled diffuse background.  It would be very interesting to test the robustness of 
this method with an in-depth study of scenarios in which the background is modelled, and we leave this for future work.

A next step would be to apply this analysis to actual {\it Fermi} LAT data.  For our mock analysis we 
have chosen an gNFW dark matter distribution, and a pulsar spatial distribution which exactly matches the resulting dark matter 
signal.  Though this choice is unrealistic, it was made to better illustrate the ability of our analysis method to distinguish 
between MSPs and DM as an origin of the GCE, in the case where the morphology alone provided minimal discriminating power.  For 
an analysis of real data, however, one should use as much information as possible.  In particular, one should use a 
better model for the pulsar distribution (see, for example, \cite{Macias:2016nev,Bartels:2017vsx,Macias:2019omb,Song:2024iup}), 
as well allow more freedom in the dark matter distribution, including allowing for 
deviations from spherical symmetry.

Especially for an analysis of actual {\it Fermi} LAT data, it is important to not only test robustness to mismodelling, 
but also to use emission models which are as accurate as possible.  These models have been improving, and with these improvements, the 
ability of methods such as the one described in this work to determine the nature of sources in the Galactic Center will improve.  The recent 
analysis from Ref.~\cite{Song:2024iup} (made public shortly after this work was completed) identified a set of spatial and 
spectral templates which currently provide the best fit to {\it Fermi} LAT data.  Our methodology can be applied using these 
updated templates, or any improved templates which are available when future analyses are performed.

Our analysis has focused entirely on constraining DM annihilation and MSP contributions to the GCE.  However, the basic methodology and SBI tools that we have developed are much more general, and present a powerful approach to extracting information from gamma-ray data in general.  One particularly promising avenue for future exploration is using similar SBI techniques to analyze the diffuse gamma-ray background (DGRB).

It is also worth noting that the approach of generating individual photons can be utilized in conjunction with a variety of 
SBI techniques, beyond the NPE method we have used here.  Although the correct modelling of the PSF which is allowed by 
using individually generated photons does not have much effect on our analysis because our pixel size is large, it may have 
an effect on future analyses involving smaller pixels.  The development of a  general-purpose tool for generating mock data 
at the level of individual photons would also be a useful topic of future work.

{\bf Acknowledgements.}
We are grateful to Kevork Abazajian and Roland Crocker for useful discussions.  
EB and KC are supported in part by NASA grant \#22-FERMI22-0011.
JK is supported in part by DOE grant DE-SC0010504. 
For facilitating portions of this research, JK wishes to acknowledge the Center for Theoretical Underground Physics and Related Areas (CETUP*), The Institute for Underground Science at Sanford Underground Research Facility (SURF), and the South Dakota Science and Technology Authority for hospitality and financial support, as well as for providing a stimulating environment.

\newpage

\appendix

\section{Alternative pulsar model}  

Here, we consider an alternative choice for the parameters of the pulsar density and luminosity distribution.  
We choose 
$n_1=0.97$, $n_2=2.60$, $L_b = 1.7 \times 10^{33} \erg / \s$, with an upper 
cutoff on $L$ at $10^{37} \erg / \s$ and a lower cutoff at $10^{30} \erg / \s$.  This benchmark 
was considered in Ref.~\cite{Bartels:2018xom} (see also~\cite{Dinsmore_2022}), and was obtained from 
a Bayesian analysis of 96 MSPs detected in gamma-ray emission by {\it Fermi} LAT.
For this model, $\langle L \rangle = 0.077 L_b$, and the number of pulsars 
(resolved or unresolved) producing the GCE is $A_{\rm core} \sim 2.9 \times 10^4$.
For this benchmark, the GCE is produced by many pulsars, each of which produces only a few gamma rays detected by 
{\it Fermi}.  As such, one expects the photon count distribution to be more Poisson-like.
The flux from core pulsars is $F 
\sim 4 \times 10^{-10} \erg~\cm^{-2} \s^{-1}$, which is near the lower end of the 
range of fluxes which could match the GCE~\citep{Dinsmore_2022}.

\begin{figure}
    \includegraphics*[width=\textwidth/2]{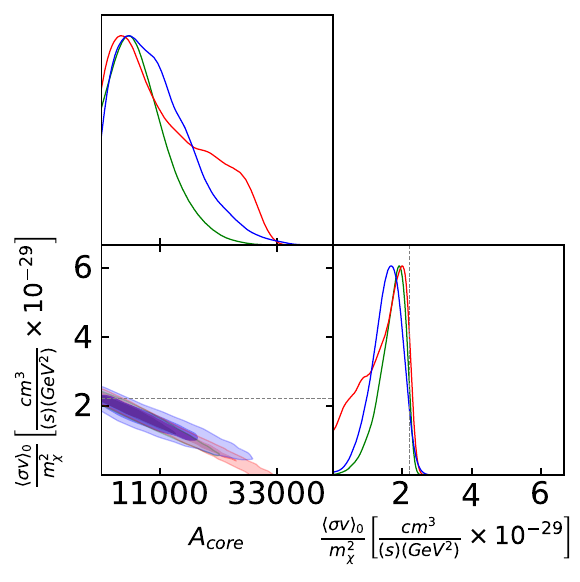}
    \includegraphics*[width=\textwidth/2]{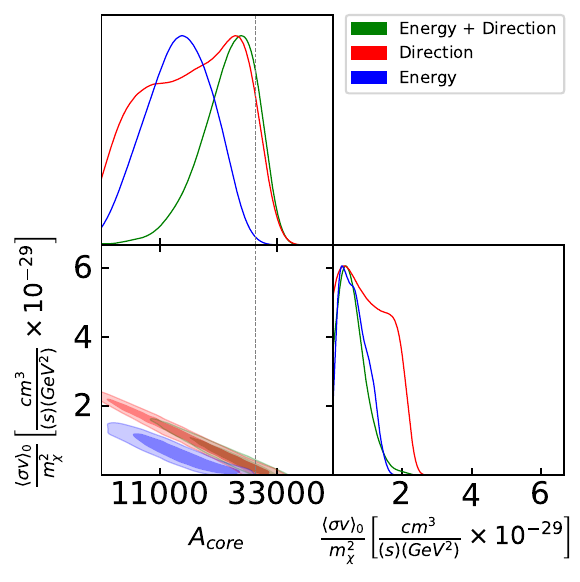}
    \caption{Similar to Fig.~\ref{fig:red}, but with a pulsar model that gives pixel-to-pixel count variations closer to a Poisson distribution.}
    \label{fig:red_p}
\end{figure}

In Figure~\ref{fig:red_p} we plot parameter constraints in the case where the GCE is due to dark matter annihilation 
exclusively (left panel) or due only to pulsars (right panel).  
Comparing to the results in Fig.~\ref{fig:red}, we see that the 
parameter constraints are somewhat weaker.  Since the number of pulsars involved is larger, non-Poisson 
fluctuations in both the counts-in-pixels and the counts-in-energy-bins are suppressed, reducing 
discriminatory power.

\bibliography{thebibVF}

\label{lastpage}
\end{document}